\documentclass[12pt]{iopart}

\usepackage{iopams}  
\usepackage{bm}
\usepackage{psfrag,graphicx,epsfig,color}
\usepackage{subfigure}
\usepackage[percent]{overpic}
\usepackage{float}
\usepackage{color}

\begin{document}

\title{A Matrix Contraction Process}

\author{Michael Wilkinson and John Grant}
\address{Department of Mathematics and Statistics, The Open University,
          Walton Hall,  Milton Keynes, MK7 6AA, England.}

\begin{abstract}

We consider a stochastic process in which independent identically distributed random matrices are multiplied and where the Lyapunov exponent of the product is positive. We continue multiplying the random matrices as long as the norm, $\epsilon$, of the product is \emph{less} than unity. If the norm is greater than unity we reset the matrix to a multiple of the identity and then continue the multiplication. We address the problem of determining the probability density function of the norm, $P_\epsilon$. We argue that,  in the limit as $\epsilon\to 0$,  $P_\epsilon\sim (\ln (1/\epsilon))^\mu \epsilon^\gamma$, where $\mu $ and $\gamma$ are two real parameters. 
 
Our motivation for analysing this \emph{matrix contraction process} is that it serves as a model for describing the fine-structure of strange attractors, where a dense concentration of trajectories results from the differential of the flow being contracting in some region. We exhibit a matrix-product model for the differential of the flow in a random velocity field, and show that there is a phase transition, with the parameter $\mu$ changing abruptly from $\mu=0$ to $\mu=-\frac{3}{2}$ as a parameter of the flow field model is varied. 
\end{abstract}

\section{Introduction}
\label{sec: 1}

Consider a random multiplicative process obtained from a sequence $a_1,a_2,a_3,\cdots$ of real positive independent identically distributed random variables, each having a finite probability of being less than unity and of being greater than unity. Let the value, $\epsilon$, of the process after $N$ steps be given by the random variable 
\begin{equation}
\label{eq: 1.1}
\epsilon= \epsilon_0 \left(\prod_{k=1}^N a_k \right)
\end{equation}
where $\epsilon_0<1$ is a positive scalar constant, providing this value is \emph{less} than unity. If this product exceeds unity we reset $\epsilon$ to the value $\epsilon_0$, and reset $N$ to 0.

This process has one of two possible types of behaviour, depending on the probability distribution of the $a_k$. The value of $\epsilon$ given by equation (\ref{eq: 1.1}) may tend to decrease as $N\to \infty$. The other possibility is that the product tends to increase and the process is repeatedly reset. We are interested in this latter case as it generates a statistically stationary sequence of values of $\epsilon$, and we can consider the probability density function (PDF), $P_\epsilon$ (throughout we use $P_X$ to denote the PDF of a random variable $X$, and $\langle X\rangle$ to denote its expectation value). 

The distribution of $\epsilon$ is most easily understood in terms of the random variable, $Z=\ln\,\epsilon$ which is a \emph{sum} of independent random numbers, $z_k=\ln\,a_k$, each having a positive mean value. In the limit as $Z\to -\infty$ a master equation for $Z$ becomes independent of $Z$, unless the probability distribution of the $z_k$ has \lq heavy tails'. This symmetry is respected by choosing a PDF of the form $P_Z=\exp(\alpha Z)$, for some coefficient $\alpha$ which must be positive to give a normalisable probability density. So, for small $\epsilon$, the PDF $P_\epsilon$, is given by a power-law of the form
\begin{equation}
\label{eq: 1.2}
P_\epsilon \sim \ \epsilon^\gamma
\end{equation}
where $\gamma=\alpha-1$. This argument showing that this \emph{scalar contraction process} has a power-law distribution for $\epsilon$ was previously presented in \cite{Gui+16}.

It is explained in \cite{Gui+16} that the scalar contraction process has applications in dynamical systems theory, because the differential of a dynamical map is, by the chain rule, a product of the differentials for each iteration. In the case of a system with one degree of freedom, involving a chaotic map of one variable, it is reasonable to model this differential as a product of independent random numbers, analogous to equation (\ref{eq: 1.1}). The resetting process addresses what happens when the separation of trajectories is no longer small and the linearisation approximation fails. Since we are concerned with small separations, we ignore the dynamics while the separation becomes large and reset the process when the separation of trajectories becomes small again.

For a system with more than one degree of freedom the differential of the dynamical map is described by a product of \emph{stability matrices}, rather than a product of scalars \cite{Ott02}. In this paper we generalise from the scalar case to consider a \emph{matrix contraction process} involving a product of $M\times M$ square matrices:
\begin{equation}
\label{eq: 1.3}
{\bf A} = \frac{\epsilon_0}{\sqrt{M}} {\bf a}_N{\bf a}_{N-1}\ldots {\bf a}_2 {\bf a}_1
\end{equation}
where the ${\bf a}_i$ are independent, identically distributed random matrices. When $\epsilon$, the (Frobenius) norm  of {\bf A}, defined by
\begin{equation}
\label{eq: 1.4}
\epsilon=\sqrt{{\rm tr}\left({\bf A}^{\rm T}{\bf A}\right)},
\end{equation}
equals or exceeds unity we reinitialise ${\bf A}$ to  $\epsilon_0/\sqrt{M}$ times
the identity matrix, so that the norm is $\epsilon_0$, and continue the iteration. This generates a sequence of values of $\epsilon$ characterized by a PDF $P_\epsilon$.  Analogy with the scalar contraction process suggests that $P_\epsilon$ is a power-law, however we claim that for these matrix-valued contraction processes a more general form may be required to describe the PDF as $\epsilon \to 0$:
\begin{equation}
\label{eq: 1.5}
P_\epsilon\sim \left(\ln\,\frac{1}{\epsilon} \right)^\mu \epsilon^\gamma
\end{equation}
where $\mu $ is another parameter.

Our objective is to justify this assertion and to show how $\gamma $ and $\mu$ can be determined. We discuss this in detail for a specific class of models, but the construction can be generalised. We find that $\gamma$ depends continuously on parameters of the model, but that $\mu$ is constant over intervals of the parameters, exhibiting discontinuous jumps. In some cases we find $\mu=0$ so that $P_\epsilon$ is a power-law.

The matrix contraction process has a natural application to describing the structure of strange attractors. In particular, it is possible to relate the distribution of $\epsilon$ to fractal dimensions. We will consider these connections in detail in a companion paper \cite{Wil+17}, where we characterise the structure of compact constellations of phase points and discuss their relation to Renyi dimensions of the attractor.

In order to simplify the discussion, we restrict ourselves to a model where ${\bf A}$ is a continuous function of some time variable, $t$, and the factors in equation (\ref{eq: 1.3}) are close to the identity so that we may write
\begin{equation}
\label{eq: 1.6}
{\bf A}(t)=\frac{\epsilon_0}{\sqrt{M}}({\bf I}+\delta {\bf a}_n)({\bf I}+\delta {\bf a}_{n-1})\cdots 
 ({\bf I}+\delta {\bf a}_2)({\bf I}+\delta {\bf a}_1)
\end{equation}
where $n \equiv {\rm Int}(t/ \delta t)$, with $\delta t$  a small increment of time. 
The elements of the $\delta {\bf a}_n$ are assumed to have the following statistics
\begin{equation}
\label{eq: 1.7}
\langle (\delta {\bf a}_n)_{ij}\rangle=0
\ ,\ \ \ 
\langle (\delta {\bf a}_n)_{ij}(\delta {\bf a}_m)_{kl}\rangle=2{\cal D}_{ijkl}\ \delta _{nm}\delta t
\end{equation}
so that the evolution of ${\bf A}(t)$ is characterised by a diffusive process, with diffusion coefficients ${\cal D}_{ijkl}$. 

We consider the simplest case  
where ${\bf A}(t)$ is a $2\times 2$ matrix. The initial value of ${\bf A}(t)$, and the value it takes whenever it is  reset, is  a scalar multiple of the $2 \times 2$ identity matrix, equal to $\frac{1}{\sqrt {2}}\epsilon_0 {\bf I}$, where $\epsilon_0$ is the initial value of the matrix norm and $0< \epsilon_0 < 1$.

\section{Evolution of the process ${\bf A}(t)$}
\label{sec: 2}

The analysis of the evolution of ${\bf A}(t)$ may be simplified by using the singular value decomposition (SVD) of ${\bf A}(t)$. This can be written in the form  (see \cite{Horn13})
\begin{equation}
\label{eq: 2.1}
{\bf A}(t)={\bf R}_1\,{\bf \Lambda}\,{\bf R}_2
\end{equation}
where ${\bf R}_1 \equiv {\bf R}(\theta_1)$ and ${\bf R}_2 \equiv {\bf R}(\theta_2)$
are rotation matrices with
\begin{equation}
\label{eq: 2.1a}
{\bf R}(\theta)=\left(\begin{array}{cc}
\cos \theta & \sin \theta \cr 
-\sin \theta & \cos \theta
\end{array}\right) \ ,
\
\end{equation}
and ${\bf \Lambda}$ is a diagonal matrix whose entires are the singular values $\lambda_1$ and $\lambda_2$:
\begin{equation}
\label{eq: 2.1b}
{\bf \Lambda}= \left( \begin{array}{cc}
									\lambda_1  &  0        \cr
									    0      & \lambda_2 
                  \end{array} \right)\ .
\end{equation}
The rotation angles and singular values depend on $t$.

Consider the process at times $n \delta t$ and $(n+1) \delta t$. Suppose that, in the time interval, the increments in matrices ${\bf R}_1$, ${\bf R}_2$ and ${\bf \Lambda}$ are, respectively,  $\delta {\bf R}_1$, $\delta {\bf R}_2$, and $\delta{\bf \Lambda}$ (we have suppressed the arguments of the matrices for clarity). Then writing ${\bf A}_n$ for ${\bf A}(n \delta t)$ we have
\begin{equation}
\label{eq: 2.1c}
{\bf A}_{n+1} = {\bf a}_{n+1}{\bf A}_{n}= \left({\bf I} + \delta {\bf a}_{n+1} \right){\bf A}_{n} 
\end{equation}
and the SVD of ${\bf A}_{n+1}$ can be written as
\begin{equation}
\label{eq: 2.1d}
{\bf A}_{n+1} = \left( {\bf R}_1 + \delta {\bf R}_1 \right ) \left( {\bf \Lambda}+
\delta{\bf \Lambda} \right) \left( {\bf R}_2 + \delta {\bf R}_2 \right) \ .
\end{equation}
Since ${\bf A}_{n}={\bf R}_1{\bf \Lambda}{\bf R}_2 $, expanding and 
comparing these expressions for ${\bf A}_{n+1}$ gives  
\begin{eqnarray}
\label{eq: 2.1e}
  \delta{\bf a}_{n+1} {\bf A}_{n} & = &
  {\bf R}_1  {\bf \Lambda} \delta{\bf R}_2  
+ {\bf R}_1 \delta{\bf \Lambda} {\bf R}_2 
+ \delta{\bf R}_1 {\bf \Lambda} {\bf R}_2 
\nonumber \\
& +& {\bf R}_1 \delta{\bf \Lambda}  \delta{\bf R}_2
+ \delta {\bf R}_1 {\bf \Lambda}  \delta{\bf R}_2
+  \delta{\bf R}_1 \delta {\bf \Lambda} {\bf R}_2 
\nonumber \\
& +& \delta{\bf R}_1 \delta {\bf \Lambda}  \delta{\bf R}_2 .
\end{eqnarray}
Pre-multiplying this equation by ${\bf R}_{1 }^{-1}$ and postmultiplying it by  
$ {\bf R}_2^{-1}{\bf \Lambda}^{-1}$  gives 
\begin{eqnarray}
\label{eq: 2.1f}
\delta \tilde{{\bf a}}_{n+1} & =& 
{\bf \Lambda} \delta{\bf R}_2 {\bf R}_2^{-1} {\bf \Lambda}^{-1}
+ \delta{\bf \Lambda}{\bf \Lambda}^{-1}
+  {\bf R}_1^{-1} \delta{\bf R}_1 
\nonumber \\
& + & \delta{\bf \Lambda}  \delta{\bf R}_2 {\bf R}_2^{-1} {\bf \Lambda}^{-1}   
+  {\bf R}_1^{-1} \delta{\bf R}_1 {\bf \Lambda} \delta{\bf R}_2  {\bf R}_2^{-1}
 {\bf \Lambda}^{-1}   
+ {\bf R}_1^{-1} \delta {\bf R}_1 \delta {\bf \Lambda}{\bf \Lambda}^{-1}
\nonumber \\
& +& {\bf R}_1^{-1} \delta{\bf R}_1  \delta {\bf \Lambda}  \delta{\bf R}_2 {\bf R}_2^{-1}{\bf \Lambda}^{-1} \ ,
\end{eqnarray}
where 
\begin{equation}
\label{eq: 2.1x}
\delta \tilde{{\bf a}}_{n+1} \equiv {{\bf R}_1}^{-1} \delta {\bf a}_{n+1} {\bf R}_1
\ .
\end{equation}
Now, in terms of the increments in the singular values, $\delta \lambda_i$, we have  
\begin{equation}
\label{eq: 2.1g}
\delta {\bf \Lambda} {\bf \Lambda}^{-1} 
=\left( \begin{array}{cc}
 \frac{\delta \lambda_1}{\lambda_1} & 0 \cr 
 0 & \frac{\delta \lambda_2}{\lambda_2}
 \end{array} \right)
= {\bf \Lambda}^{-1} \delta {\bf \Lambda}  
\end{equation}
and, to second order in the  $\delta \theta_i$,
\begin{equation}
\label{eq: 2.1h}
{\bf R}_{i}^{-1} \delta {\bf R}_{i}=
\left( 
	\begin{array}{cc}
		-\frac{\delta \theta_i^{2}}{2} & \delta \theta_i \cr 
		-\delta \theta_i & -\frac{\delta \theta_i^{2}}{2} \cr
\end{array} 
\right)
=\delta {\bf R}_{i} {\bf R}_{i}^{-1} \ .
\end{equation}
Therefore, to the second order of increments, equation (\ref{eq: 2.1f}) gives  
\begin{equation}
\label{eq: 2.1i}
\!\!\!\!\!\!\!\!\!\!\!\!\!\!\!\!\!\!\!\!\!\!\!\!\!\!\!\!\!\!\!\!\!\!\
\delta \widetilde{{\bf a}}_{n+1} =
\left(
\begin{array}{cc}
\frac{\delta \lambda_1}{\lambda_1} - \frac{1}{2} \left(\delta \theta_1^{2} 
+ \delta \theta_2^{2} \right) - \left( \frac{\lambda_2}{\lambda_1} \right) \delta \theta_1 \delta \theta_2 \ , 
& 
\left(1+\frac{\delta \lambda_2}{\lambda_2} \right)\delta \theta_1
+ \left( \frac{\lambda_1}{\lambda_2} \right) \left(1+\frac{\delta \lambda_1}{\lambda_1} \right)
\delta \theta_2  \ ,
\cr 
-\left( 1+\frac{\delta \lambda_1}{\lambda_1} \right)\delta \theta_1 
-\left( \frac{\lambda_2}{\lambda_1} \right) \left( 1+\frac{\delta \lambda_2}{\lambda_2} \right) 
\delta \theta_2\ ,
&
\frac{\delta \lambda_2}{\lambda_2} -  \frac{1}{2} \left( \delta \theta_1^{2} 
+ \delta \theta_2^{2} \right) - \left( \frac{\lambda_1}{\lambda_2} \right) \delta \theta_1 \delta \theta_2 
\cr
\end{array}
\right)
\end{equation}
In the limit as $\delta t \to 0$, equation (\ref{eq: 2.1i}) reduces to a system of coupled stochastic differential equations (SDEs) for the singular values and rotation angles: 
\begin{eqnarray}
\label{eq: 2.1j}
\frac{\rm{d} \lambda_1}{\lambda_1}  &=& \rm{d}{\tilde a}_{11} 
+ \frac{1}{2} \left(\rm{d} \theta_1^{2} + \rm{d} \theta_2^{2} \right) + \nu \rm{d} \theta_1\rm{d} \theta_2
\\
\cr
\label{eq: 2.1k}
\frac{\rm{d} \lambda_2}{\lambda_2} &=& \rm{d}{\tilde a}_{22}  + \frac{1}{2} \left(\rm{d} \theta_1^{2} 
+ \rm{d} \theta_2^{2} \right) + \left( \frac{1}{\nu} \right) \rm{d} \theta_1\rm{d} \theta_2  
\\
\label{eq: 2.1l}
\rm{d} \theta_1 &=& - \frac{ \frac{1}{\nu}\left(1+\frac{\rm{d} \lambda_1}{\lambda_1} \right)
\rm{d}{\tilde a}_{21}+{\nu}\left(1 + \frac{\rm{d} \lambda_2}{\lambda_2}\right)\rm{d}{\tilde a}_{12} } 
{\frac{1}{\nu}\left(1+\frac{\rm{d} \lambda_1}{\lambda_1} \right)^2 - {\nu}
\left(1 + \frac{\rm{d} \lambda_2}{\lambda_2} \right)^2 }
\\
\label{eq: 2.1m}
\rm{d} \theta_2 &=& \frac{ \left(1 + \frac{\rm{d} \lambda_1}{\lambda_1}\right)\rm{d}{\tilde a}_{12} 
+\left(1+\frac{\rm{d} \lambda_2}{\lambda_2} \right) \rm{d}{\tilde a}_{21}}
{\frac{1}{\nu} \left(1 + \frac{\rm{d} \lambda_1}{\lambda_1}\right)^2-\nu
\left(1+\frac{\rm{d} \lambda_2}{\lambda_2} \right)^2 }
\end{eqnarray}
where $ \nu \equiv \lambda_2/\lambda_1 $ and we have suppressed the time subscript on the $ \rm{d}{\tilde a}_{ij} $ (the remaining subscripts denote the row and column position of the matrix element). It is convenient to replace the singular values with logarithmic variables ${Z_i = \rm ln}\lambda_i$: we have
\begin{equation}
\label{eq: 2.2}
{\rm d}Z_i = \frac{{\rm d} \lambda_i}{\lambda_i}-\frac{1}{2} \left( \frac{{\rm d} 
\lambda_i}{\lambda_i} \right)^2
\end{equation}
then, retaining only terms upto the second order in small increments, we obtain the following set of SDEs for the increments of the $Z_i$ and $\theta_i$ in terms of the matrix elements, ${\rm d}{\tilde a}_{ij}$. Defining
\begin{equation}
\label{eq: 2.6b}
\alpha \equiv \frac{\nu}{\left( \nu^2-1 \right)} 
\end{equation} 
we find:
\begin{eqnarray}
\label{eq: 2.3}
\rm{d} Z_1&=& \rm{d}\tilde{a}_{11} 
-\frac{1}{2}\rm{d}\tilde{a}_{11}^2
-\frac{\alpha}{2}\left[\frac{1}{\nu} \rm{d}\tilde{a}_{21}^2
+2\nu\rm{d}\tilde{a}_{12}\rm{d}\tilde{a}_{21}
+\nu \rm{d}\tilde{a}_{12}^2\right]
\\
\label{eq: 2.4}
\rm{d}Z_2 &=& \rm{d}\tilde{a}_{22} 
-\frac{1}{2}\rm{d}\tilde{a}_{22}^2
+\frac{\alpha}{2}\left[\nu\rm{d}\tilde{a}_{12}^2
+\frac{2}{\nu}\rm{d}\tilde{a}_{12}\rm{d}\tilde{a}_{21}
+\frac{1}{\nu}\rm{d}\tilde{a}_{21}^2\right]
\end{eqnarray}
and 
\begin{eqnarray}
\label{eq: 2.5}
\rm{d}\theta_1 &=& \alpha\left[\nu \rm{d}\tilde{a}_{12}+\frac{1}{\nu}\rm{d}\tilde{a}_{21}\right]
+\alpha^2\rm{d}\tilde{a}_{11}\left[2\rm{d}\tilde{a}_{12}
+\frac{\nu^2+1}{\nu^2}\rm{d}\tilde{a}_{21}\right]
\nonumber \\
&&-\alpha^2\rm{d}\tilde{a}_{22}\left[2\rm{d}\tilde{a}_{21}
+(\nu^2+1)\rm{d}\tilde{a}_{12}\right]
\\
\label{eq: 2.6}
\rm{d}\theta_2 &=& 
-\alpha \left[\rm{d}\tilde{a}_{12}+\rm{d}\tilde{a}_{21}\right]
-\frac{\alpha^2}{\nu}\rm{d}\tilde{a}_{11}\left[2\rm{d}\tilde{a}_{12}
+(\nu^2+1)\rm{d}\tilde{a}_{21}\right]
\nonumber \\
&&+\alpha^2\rm{d}\tilde{a}_{22}\left[2\nu \rm{d}\tilde{a}_{12}
+\frac{(\nu^2+1)}{\nu} \rm{d}\tilde{a}_{21}\right]
\ .
\end{eqnarray}
These equations can now be used to produce a Fokker-Planck equation for the joint probability density of the variables, $Z_i$ and $\theta_i$. The same Fokker-Planck equation arises if we replace the second-order terms by their mean values so we simplify the equations by taking expectation values of the second-order terms. At this stage there is nothing to distinguish between $\lambda_1$ and $\lambda_2$. However we do expect that both $\lambda_1$ and $\lambda_2$ have non-zero and distinct Lyapunov exponents, so that either $\nu\to 0$ or $\nu\to\infty$ as $t\to \infty$, with probability unity. Which case occurs is random and equiprobable. Let us assume that symmetry breaks so that $\nu\to 0$ in the long time limit. Accordingly, we consider this limit (noting that $\alpha/\nu \to -1$ as $\nu\to 0$). We obtain the following Langevin equations:
\begin{eqnarray}
\label{eq: 2.7}
\rm{d} Z_1=& \rm{d}\tilde{a}_{11} 
-\frac{1}{2}\langle \rm{d}\tilde{a}_{11}^2\rangle
+\frac{1}{2}\langle\rm{d}\tilde{a}_{21}^2\rangle
\\
\cr
\label{eq: 2.8}
\rm{d}Z_2 =&\rm{d}\tilde{a}_{22} 
-\frac{1}{2}\langle \rm{d}\tilde{a}_{22}^2\rangle
-\langle \rm{d}\tilde{a}_{12}\rm{d}\tilde{a}_{21}\rangle
-\frac{1}{2}\langle \rm{d}\tilde{a}_{21}^2\rangle
\\
\cr
\label{eq: 2.9}
\rm{d}\theta_1 =&-\rm{d}\tilde{a}_{21} +\langle \rm{d}\tilde{a}_{11}\rm{d}\tilde{a}_{21}\rangle
\\
\cr
\label{eq: 2.10}
\rm{d}\theta_2 =& 0
\ .
\end{eqnarray}
Note that $\theta_2$ freezes as $t\to \infty$. This is to be expected because the direction along which the norm is most rapidly increasing is expected to approach a limit as $t\to \infty$. Equations (\ref{eq: 2.7}) and (\ref{eq: 2.8}) are independent of the variables $Z_i$. These quantities therefore have a diffusive evolution at long times and we have 
\begin{equation}
\label{eq: 2.11}
\langle Z_i(t)\rangle =v_i t
\ ,\ \ \ 
\langle (Z_i-v_it)(Z_j-v_jt)\rangle =2{\cal D}_{ij}t 
\ .
\end{equation}
The joint probability density of the $Z_i$ after time $t$ is therefore determined using a Green's 
function
\begin{equation}
\label{eq: 2.12}
G(Z_1,Z_2,t)=\frac{1}{4\pi\sqrt{{\rm det}({\bf D})}t}
\exp\left[-S(\mbox{\boldmath$Z$},t)\right]
\end{equation}
where 
\begin{equation}
\label{eq: 2.13}
S(\mbox{\boldmath$Z$},t)=
\frac{1}{4t}(\mbox{\boldmath$Z$}-\mbox{\boldmath$v$}t)
\cdot{\bf D}^{-1} (\mbox{\boldmath$Z$}-\mbox{\boldmath$v$}t)
\ .
\end{equation}
Because we have assumed that $\nu \to 0$, these equations are valid provided that $Z_1-Z_2$ is sufficiently large.

\section{Matrix contraction}
\label{sec: 3}

\subsection{Principles of calculation}
\label{sec: 3.1}

Our objective is to understand the distribution of the norm, $\epsilon$, for the  matrix contraction process. In terms of the singular values and the logarithmic variables the norm of the matrix ${\bf A}(t)$ is  
\begin{equation}
\label{eq: 3.1}
\epsilon=\sqrt{\lambda_1^2+\lambda_2^2}=\sqrt{(\exp(2Z_1)+\exp(2Z_2})
\ . 
\end{equation}
The PDF of $\epsilon$ may therefore be obtained from the joint PDF, $P_{(Z_1,Z_2)}$, of $Z_1$ and $Z_2$. 
In section \ref{sec: 2} we showed that $\mbox{\boldmath$Z$}=(Z_1,Z_2)$ undergoes a diffusive process with drift. The matrix contraction process may therefore be represented by a point, or a notional particle, in the $(Z_1,Z_2)$ plane which undergoes advective diffusion, with drift velocity $\mbox{\boldmath$v$}$ and diffusion tensor ${\bf D}$. 

\begin{figure*}[!ht]
\begin{center}
  \begin{overpic}[width=0.70\textwidth, grid=false, tics=20]{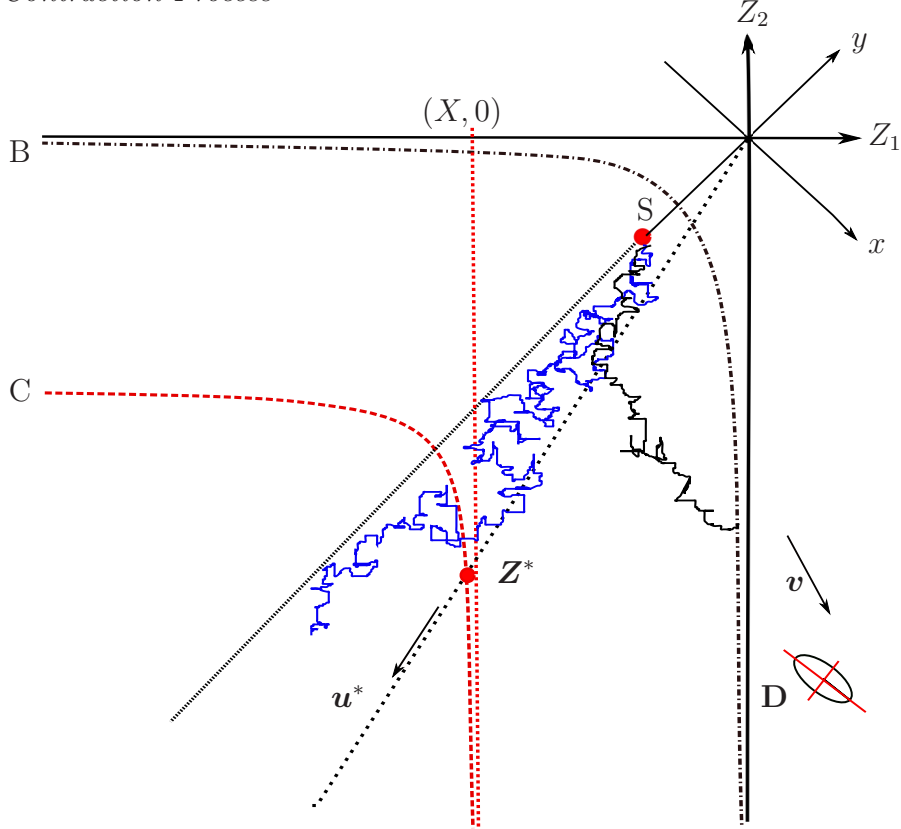} 
    \put(100,83){$Z_1$}
    \put(84,98){$Z_2$}
		\put(98,95){$y$}
    \put(100,70){$x$}
		\put(46,86){$(X,0)$}
		\put(72,74){${\rm S}$}
		\put(-4,81){${\rm B}$}
	  \put(-4,52){${\rm C}$}
    \put(55,30){$\mbox{\boldmath$Z$}^{\ast}$}
    \put(90,29){$\mbox{\boldmath$v$}$}
    \put(87,15){${\bf D}$}
		\put(35,15){$\mbox{\boldmath$u$}^{\ast}$}	  
  \end{overpic} 
\end{center}
\caption{\label{fig: 1} (Colour online). 
The matrix contraction process is represented by an ensemble of particles which undergo advective diffusion in the $Z_1Z_2$ plane, with drift velocity $\mbox{\boldmath$v$}$ and diffusion tensor ${\bf D}$. The resetting operation corresponds to the diffusing particles being absorbed on the boundary B, and replaced by new particles at the source point S. The line $Z_1=Z_2$ is a reflecting barrier. The curve C is a contour of the matrix norm, $\epsilon$. Two representative paths are shown: the blue path shows a particle which is reflected and the black path shows a particle which is absorbed.
}
\end{figure*}

This process is illustrated schematically in figure \ref{fig: 1}. Since we have ordered the singular values so that $Z_1\ge Z_2$, the line $Z_1=Z_2$ is a reflecting boundary. The resetting process occurs when the norm of the matrix is equal to unity, which is represented by a contour $\rm{B}$. The contour ${\rm B}$ is thereforean absorbing boundary. When the representative point reaches this absorbing boundary, $\bf A(t)$ is re-set to $\frac{1}{\sqrt{2}}\epsilon_0 {\mathbf{I}}$ and the representative diffusing \lq particle' is re-introduced at the source at the point ${\rm S}=\left( {\rm ln} (\epsilon_0/\sqrt{2}),{\rm ln}(\epsilon_0/\sqrt{2}) \right)$. The process may therefore be modelled by an ensemble of \lq particles', which diffuse and drift in a wedge-like domain of the $(Z_1,Z_2)$ plane, being reflected at one edge and absorbed at the other. The loss of particles by absorption on ${\rm B}$ is balanced by particles being injected at ${\rm S}$.

Consider how to determine the probability density $P_{(Z_1,Z_2)}$ for the  representative point to reach $(Z_1,Z_2)$. Let $p(Z_1,Z_2,t,\Delta t)$ be  the probability density for the point to be at $(Z_1,Z_2)$ at time $t$ after it is emiitted from S, and to be absorbed on B after time $t+\Delta t$. The probability density $P_{(Z_1,Z_2)}$ is obtained by integrating over the times:
\begin{equation}
\label{eq: 3.1a}
P_{(Z_1,Z_2)}=\int_0^\infty {\rm d}t\int_0^\infty {\rm d}\Delta t\ p(Z_1,Z_2,t,\Delta t)
\end{equation}
Because the diffusion process is Markovian, we can write $p$ as a product of two functions:  $p(Z_1,Z_2,t,\Delta t)=G(Z_1,Z_2,t)p_2(Z_1,Z_2,\Delta t)$, where $G(Z_1,Z_2,t)$ is the Green's function (probability density to reach $(Z_1,Z_2)$ after time $t$), and $p_2(Z_1,Z_2,\Delta t)$ is the probability density for a particle released at $(Z_1,Z_2)$ to be absorbed on B after time $\Delta t$. From this definition of $p_2$, we have 
\begin{equation}
\label{eq: 3.1b}
\int_0^\infty {\rm d}\Delta t\ p_2(Z_1,Z_2,\Delta t)=1
\end{equation}
so that 
\begin{equation}
\label{eq: 3.2}
P_{(Z_1,Z_2)}=\int_0^\infty {\rm d}t\ G(Z_1,Z_2,t)
\ .
\end{equation}
Strictly speaking, $G(Z_1,Z_2,t)$ should be the Green's function for reaching $(Z_1,Z_2)$ after time $t$ by a path that does not cross B. However, when we determine the probability density of $\epsilon$, we find that we require $P_{(Z_1,Z_2)}$ for positions which are not close to the absorbing boundary B. In this case we can use the Green's function for the non-absorbing boundary. In cases where the required values of ${(Z_1,Z_2)}$ are not close to the reflecting line, we can use  equation (\ref{eq: 2.12}) to approximate the Green's function.

Following the approach used in \cite{Gra+15}, we use the Laplace principle to estimate the integral in equation (\ref{eq: 3.2}). At the time $t^\ast$ when the propagator is maximal,  $\partial S/\partial t(\mbox{\boldmath$Z$},t^\ast)=0$, so that $t^\ast$ is given by 
\begin{equation}
\label{eq: 3.2a}
t^\ast=
\sqrt{\frac
{\mbox{\boldmath$Z$}\cdot {\bf D}^{-1}\mbox{\boldmath$Z$}}
{\mbox{\boldmath$v$}\cdot {\bf D}^{-1}\mbox{\boldmath$v$}}
}
\end{equation}
and, therefore 
\begin{equation}
\label{eq: 3.3}
P_{(Z_1,Z_2)}\sim \exp\left[-\Phi(\mbox{\boldmath$Z$})\right]
\end{equation}
where 
\begin{equation}
\label{eq: 3.4}
\Phi (\mbox{\boldmath$Z$})=S(\mbox{\boldmath$Z$},t^\ast)  =\frac{1}{2}\left[\sqrt{\mbox{\boldmath$Z$}\
\cdot{\bf D}^{-1}\mbox{\boldmath$Z$}}\sqrt{\mbox{\boldmath$v$}\
\cdot{\bf D}^{-1}\mbox{\boldmath$v$}}
-\mbox{\boldmath$Z$}\cdot{\bf D}^{-1}\mbox{\boldmath$v$}
\right]
\ .
\end{equation}
The function $\Phi (Z_1,Z_2)$ may be interpreted as the height, above the $(Z_1,Z_2)$ plane, of a tilted conical surface which touches this plane along the ray $\mbox{\boldmath$Z$}=\lambda \mbox{\boldmath$v$}$, where $\lambda$ is a positive real parameter. Since this cone touches the plane in the direction of the drift vector $\mbox{\boldmath$v$}$, \lq downwind' of the origin, along any other ray through the origin $\Phi (Z_1,Z_2)$ must increase linearly with distance from the origin. Therefore, asymptotically, $P_{(Z_1,Z_2)}$ decays exponentially with distance from the origin along such a ray.   
We remark that this construction for $P_{(Z_1,Z_2)}$ is similar in structure to the result obtained in \cite{Wil+15}.

Recall that we wish to determine the probability density for the matrix norm, $\epsilon$, to reach a very small value. Accordingly, we consider the form of lines of constant $\epsilon$ in the $(Z_1,Z_2)$ plane. These lines are determined from equation (\ref{eq: 3.1}), together with the condition that $Z_1\ge Z_2$. The contour ${\rm C}$ in figure \ref{fig: 1} is one such line. 

It is clear from the form of (\ref{eq: 3.1}) that the contours of constant $\epsilon$ are asymptotic to the vertical line $Z_1=\ln\,\epsilon$. The probability density to reach $\epsilon$ can therefore be
obtained from that of $X={\rm ln}\ \epsilon$, which can be estimated using the expression (\ref{eq: 3.3}) on the segment of the asymptote lying on or below the line $Z_1=Z_2$.

The PDF of $X$ is obtained by applying Laplace's principle again, so that we determine the value $\mbox{\boldmath$Z$}^\ast$ that minimises $\Phi(Z_1,Z_2)$ on the line $(Z_1,Z_2)=(X,X-Y)$, with $0 \le Y<\infty$. Because $\Phi(Z_1,Z_2)$ increases linearly along any ray, we find that the saddle point $\mbox{\boldmath$Z$}^{\ast}$, at which $\Phi (\mbox{\boldmath$Z$})$ has a minimum value, lies along the \emph{direction}, $\mbox{\boldmath$u$}^\ast$ which minimises (\ref{eq: 3.4}) on the line $(Z_1,Z_2)=-(1,\eta)$, with $\eta\ge 1$. If the minimum lies at $\eta^\ast$, we have ${\bm u}^\ast=-(1,\eta^\ast)$. 
Since $\mbox{\boldmath$Z$}^\ast = |X| \mbox{\boldmath$u$}^\ast$, the probability density of $X$ is therefore of the form
\begin{equation}
\label{eq: 3.6}
P_X \sim \exp[-\Phi(\mbox{\boldmath$u$}^\ast)|X|]
\ .
\end{equation}
This is consistent with $P_\epsilon$ having a power-law distribution, and suggests that the exponent $\gamma$ in (\ref{eq: 1.5}) is
\begin{equation}
\label{eq: 3.7}
\gamma=[\Phi(\mbox{\boldmath$u$}^\ast)-1]
\ .
\end{equation}
In the discussion above it has been assumed that the saddle point $\mbox{\boldmath$Z$}^{\ast}$,  lies below the line $Z_1=Z_2$, so that $\eta^\ast>1$, as illustrated schematically in figure \ref{fig: 1}. We refer to this case as the \emph{non-degenerate case}. However, the minimum of $\Phi(Z_1,Z_2)$ 
along a line of constant $Z_1$ may occur at a \emph{physically inaccesible} point, i.e. one for which $Z_2 \geq Z_1$ or, equivalently, $\eta^\ast \leq 1$. In this case, which we refer to as the \emph{degenerate} 
case, the discussion in section (\ref{sec: 3}) must be replaced by a consideration of what happens in the neighbourhood of the boundary point $(Z_1=X,Z_2=X)$. This degenerate case is considered in section \ref{sec: 4}.

\subsection{Estimate for the pre-exponential factor}
\label{sec: 3.2}

We have argued that $P_X\sim \exp(\alpha |X|)$, with $\alpha =-\Phi(\mbox{\boldmath$u$}^\ast)$, 
where $\mbox{\boldmath$u$}^\ast$ minimises $\Phi$ along the line $Z_1=X$.
The exponential form of $P_X$ is consistent with a power-law, $P_\epsilon\sim \epsilon^\gamma$, with 
$\gamma=-\left(\alpha+1\right)$, 
but it is necessary to examine the pre-exponential factor to determine the true asymptotic form of $P_\epsilon$ as $\epsilon\to 0$. To this end, we consider in more detail the Laplace estimates of the integrals. 

First consider the integration over $t$: the Laplace method applied to (\ref{eq: 3.2}) yields
\begin{equation}
\label{eq: 3.8}
P_{(Z_1,Z_2)}\sim \frac{\exp[-\Phi(\mbox{\boldmath$Z$})]}
{t^\ast\sqrt{2\pi{\rm det}({\bf D})}}\left(\frac{\partial^2S}{\partial t^2}
(\mbox{\boldmath$Z$},t^\ast)\right)^{-1/2}
\ .
\end{equation}
The stationary point is at $t^{\ast}$, given by equation (\ref{eq: 3.2a}) and we find
\begin{equation}
\label{eq: 3.10}
\frac{\partial^2 S}{\partial t^2}(\mbox{\boldmath$Z$},t^\ast)
=\frac{(\mbox{\boldmath$v$}\cdot {\bf D}^{-1}\mbox{\boldmath$v$})^{3/2}}
{2(\mbox{\boldmath$Z$}\cdot {\bf D}^{-1}\mbox{\boldmath$Z$})^{1/2}}
\ .
\end{equation}
It follows that
\begin{equation}
\label{eq: 3.11}
P_{(Z_1,Z_2)}\sim K (\mbox{\boldmath$Z$}\cdot {\bf D}^{-1}\mbox{\boldmath$Z$})^{-1/4}
\exp[-\Phi(\mbox{\boldmath$Z$})]
\end{equation}
where $K$ is independent of $\mbox{\boldmath$Z$}$.

Now, in order to determine $P_X$ we integrate (\ref{eq: 3.11}) down the asymptote $Z_1=X$ from the point $\left(X,X\right)$, i.e. over $Z_2$ with $Z_2< X$, using the Laplace approximation for the second time. We are concerned only with the way in which $P_X$  depends on $Z_1$, and not on the precise form of any coefficients. We can therefore avoid detailed calculation of coefficients by using power-counting arguments. The expression for the second derivative of $\Phi(Z_1,Z_2)$ with respect to $Z_2$ contains  terms proportional to $({\bm Z}\cdot {\bm D}^{-1}{\bm Z})^{-1/2}$, and other terms which scale as $Z_1^{-1}$. Performing the Gaussian integral therefore introduces a factor $Z_1^{1/2}$  which cancels the $Z_1$ dependence of the factor $({\bm Z}\cdot {\bm D}^{-1}{\bm Z})^{-1/4}$ in (\ref{eq: 3.11}). At leading order, there is therefore no overall $Z_1$ dependence in the coefficient of the exponential term in the expression for  $P_X$ and we conclude that $P_X\sim \exp[-\Phi(\mbox{\boldmath$u$}^\ast) |X|]$. Therefore in the non-degenerate case the probability density $P_\epsilon$ is of the form (\ref{eq: 1.5}) with 
\begin{equation}
\label{eq: 3.12}
\gamma=[\Phi(\mbox{\boldmath$u$}^\ast)-1]
\ ,\ \ \ 
\mu=0
\ .
\end{equation}

\section{Treating the degenerate case}
\label{sec: 4}

If there is no stationary point of $\Phi (\mbox{\boldmath$Z$})$ with $Z_2<Z_1$, 
then the discussion in section \ref{sec: 3.2} does not apply. 
We now consider what happens in this degenerate case.

\subsection{Making a coordinate transformation}
\label{sec: 4.1}

We have seen that the matrix multiplication process in our model corresponds 
to the diffusive evolution of the variables $Z_1$ and $Z_2$. From equations (\ref{eq: 1.7}) it follows
that the diffusion tensor is a symmetric $2\times 2$ matrix with
equal diagonal elements, it can therefore be diagonalised by rotating the coordinate axes through $\pi/4$. We therefore find it convenient to consider the advective diffusion process in the coordinate system $x=(Z_1-Z_2)$ and $y=(Z_1+Z_2)$, where the diffusion tensor is diagonal (see figure \ref{fig: 1}).

In the  $x$, $y$ coordinate system, the source is at $\left( 0, {\rm ln} \left( \epsilon_0^2/2 \right ) \right) $, the $y$ axis is a reflecting barrier and the absorbing boundary is approximated by the line $y=-x$. 
The equations of motion, (\ref{eq: 2.5}), transform into
\begin{eqnarray}
\label{eq: 4.1.1}
{\rm d}x&=&\left({\rm d}\tilde{a}_{11}-{\rm d}\tilde{a}_{22}\right) 
 + \frac{1}{2}\left(\langle{\rm d}\tilde a_{22}^2 \rangle-
\langle {\rm d} \tilde a_{11}^2 \rangle \right)
\nonumber \\
&& +\frac{\nu}{1-\nu^2}\left[\nu \langle {\rm d}\tilde{a}_{12}^2 \rangle +\frac{1}{\nu} \langle {\rm d}a_{21}^2 \rangle
+ \left(\nu+\frac{1}{\nu}\right) \langle {\rm d}\tilde{a}_{12}{\rm d}\tilde{a}_{21} \rangle \right]
\nonumber \\
{\rm d}y &=&\left({\rm d}\tilde{a}_{11}+{\rm d}\tilde{a}_{22}\right)-
\frac{1}{2}\left( \langle{\rm d}\tilde{a}_{11}^2 \rangle+ \langle{\rm d}\tilde{a}_{22}^2 \rangle \right)
- \langle {\rm d}\tilde{a}_{12}{\rm d}\tilde{a}_{21} \rangle
\ .
\end{eqnarray}
From these it follows that the drift velocity in the $y$ direction, $v_y$, is a constant and the drift velocity in the $x$ direction, $v_x$, is a function of $\nu=\exp(-x)$ and therefore of $x$. Also, 
 the diffusion tensor for the fluctuations of $x$ and $y$ is diagonal, with diffusion coefficients 
 $D_x$, $D_y$, which are independent of $x$ and $y$.

Therefore the dynamics in the $y$ direction is simple: diffusion with a constant drift velocity. The dynamics in the $x$ direction is more complex: diffusion with a drift velocity which is a function of $x$.

The motion in the $x$ and $y$ directions are independent. In section \ref{sec: 4.2} we consider motion in the $x$ direction alone, in order to describe the effect of the reflecting boundary at $x=0$. In section \ref{sec: 4.3} we combine the results for the $x$ and $y$ motions to model the distribution of $\ln \epsilon$.

\subsection{One-dimensional diffusion with reflecting wall}
\label{sec: 4.2}

The $x$-coordinate of the particle representing the matrix contraction process undergoes diffusion with a constant diffusion coefficient $D_x$, and drift with a position-dependent drift velocity, which is a function of $x$ alone. This velocity, $v_x(x)$, approaches a positive constant value, $v_0$, as $x\to\infty$, and it approaches $\infty$ as $x\to 0$, so that $x=0$ is a reflective barrier.  We shall assume that the particle is released at $x=x_0$ at time $t=0$ and we wish to determine the distribution $P_x(x,t)$ at later times.

The probability density satisfies 
\begin{equation}
\label{eq: 4.2.1}
\frac{\partial P_x}{\partial t}=D_x\frac{\partial^2P_x}{\partial x^2}
-\frac{\partial}{\partial x}[v_x(x)P_x]
\ .
\end{equation}
This equation can be transformed to a Hermitean form by writing
\begin{equation}
\label{eq: 4.2.2}
P_x=\exp[\chi(x)]\psi(x,t)
\ ,\ \ \ 
\chi(x)=\frac{1}{2D_x}\int^x{\rm d}x'\ v_x(x')
\ .
\end{equation}
The function $\psi(x,t)$ satisfies a Schr\"odinger-like equation, with a Hermitian operator 
$\hat{\cal H}$:
\begin{equation}
\label{eq: 4.2.3}
\frac{\partial\psi}{\partial t}=D_x\frac{\partial^2\psi}{\partial x^2}-V(x)\psi 
\equiv-\hat{\cal H}\psi
\end{equation}
where
\begin{equation}
\label{eq: 4.2.4}
V(x)=\frac{1}{2}v_x^{\prime}(x)+\frac{[v_x(x)]^2}{4D_x}
\ .
\end{equation}
By introducing a nominal absorbing barrier at x=L we can develop the  solution to equation (\ref{eq: 4.2.1}) in the finite interval [0,L] as an infinite series of orthonormal eigenfunctions of the Hermitean operator $\hat{\cal H}$;  we then choose $L$ sufficiently large that its value may be assumed to have no influence. The details of this approach may be found in section 7.1 of \cite{Mah09}; we merely quote the result:
\begin{equation}
\label{eq: 4.2.11}
\!\!\!\!\!\!\!\!\!\!\!\!\!\!\!\!\!\!\!\!\!\!\!\!\!\!\!\!\!\!\!\!\!\!\!\!\!\!\
P_x(x,t)=\frac{2}{L}\exp\left(\frac{v_0(x-x_0)}{2D_x}-\frac{v_0^2t}{4D_x}\right)
\sum_{n=1}^\infty\exp\left(-\frac{n^2\pi^2D_xt}{L^2}\right)
\sin\left(\frac{n\pi x}{L}\right)\sin\left(\frac{n\pi x_0}{L}\right)
\ .
\end{equation}
Approximating the sum by an integral gives
\begin{eqnarray}
\label{eq: 4.2.12}
P_x(x,t)&=&\frac{1}{L}\exp\left(\frac{v_0(x-x_0)}{2D_x}-\frac{v_0^2t}{4D_x}\right) 
\int_0^\infty {\rm d}n\ \exp\left(-\frac{n^2\pi^2D_xt}{L^2}\right)
\nonumber \\
 &\times& \left[\cos\left(\frac{n\pi (x-x_0)}{L}\right)-\cos\left(\frac{n\pi (x+x_0)}{L}\right)\right] \ .
\end{eqnarray}
Using the standard integral
\begin{equation}
\label{eq: 4.2.13}
\int_{-\infty}^\infty{\rm d}x\ \exp(-\alpha x^2)\cos(kx)=\sqrt{\frac{\pi}{\alpha}}\exp(-k^2/4\alpha)
\end{equation}
we obtain the solution 
\begin{eqnarray}
\label{eq: 4.2.14}
P_x(x,t)&=&\frac{1}{\sqrt{4\pi D_xt}}\exp\left(\frac{v_0(x-x_0)}{2D_x}-\frac{v_0^2t}{4D_x}\right)
\nonumber \\
&&\times \left[\exp\left(-\frac{(x-x_0)^2}{4D_xt}\right)
-\exp\left(-\frac{(x+x_0)^2}{4D_xt}\right)\right]
\ .
\end{eqnarray}
In the limit where $x_0\to 0$, we have
\begin{equation}
\label{eq: 4.2.15}
P_x(x,t)\sim \frac{x_0}{\sqrt{4\pi D_x^3}}\frac{x}{t^{3/2}}\exp\left(\frac{v_0x}{2D_x}\right)
\exp\left(-\frac{x^2}{4D_xt}\right)
\exp\left(-\frac{v_0^2t}{4D_x}\right)
\ .
\end{equation}

\subsection{Implications for degenerate case}
\label{sec: 4.3}

We are interested in the probability density for the representative particle to reach the line in $(Z_1,Z_2)$ space corresponding to a given value of $X=\ln \epsilon$. In terms of the variables $x=(Z_1-Z_2)$ and $y=(Z_1+Z_2)$ we have
\begin{equation}
\label{eq: 4.3.0}
X\equiv g(x,y)=\ln \sqrt{\exp(y+x)+\exp(y-x)}=\frac{y}{2}+\frac{1}{2}\ln \left(2\cosh\, x\right)
\end{equation}
The probability density for $X$ is
\begin{eqnarray}
\label{eq: 3.4.01}
P_X&=&\int_0^\infty{\rm d}x\int_0^\infty{\rm d}y\ 
\delta \left(X-g(x,y)\right)\,P_{(x,y)}(x,y)
\nonumber \\
&=&2\int_0^\infty{\rm d}x\int_0^\infty{\rm d}y\ 
\delta \left(y-2X+\ln (2\cosh\,x)\right)\,P_{(x,y)}(x,y)
\nonumber \\
&=&2\int_0^\infty{\rm d}x\ P_{(x,y)}(x,2X-\ln(2\cosh\,x))
\ .
\end{eqnarray} 
Because the diffusive motions in the $x$ and $y$ coordinates are independent, the probability density to reach $(x,y)$ after time $t$ is expressed as a product:
\begin{equation}
\label{eq: 4.3.2}
G(x,y,t)=P_x(x,t)P_y(y,t)
\ .
\end{equation}
The probability density $P_y(y,t)$ is that for a simple advection-diffusion process:
\begin{equation}
\label{eq: 4.3.3}
P_y(y,t)=\frac{1}{\sqrt{4\pi D_y t}}\exp\left[-\frac{(y-v_yt)^2}{4D_yt}\right]
\ .
\end{equation}
The probability density $P_x(x,t)$, which must take account of the fact that $x=0$ is a reflecting barrier, is given by equation (\ref{eq: 4.2.15}). Therefore
\begin{eqnarray}
\label{eq: 4.3.4}
P_X&=&2\int_0^\infty {\rm d}x\int_0^\infty{\rm d}t\ P_x(x,t)P_y(2X-\ln(2\cosh\,x),t)
\nonumber \\
&=&\frac{x_0}{2\pi D_x\sqrt{D_xD_y}}\exp\left(\frac{X v_y}{D_y}\right)
\int_0^\infty {\rm d}x\ x \exp\left(\frac{v_0x}{2D_x}-\frac{\ln(2\cosh\, x)v_y}{2D_y}\right)
\nonumber \\
&&\times
I(x,2X-\ln(2\cosh\, x))
\end{eqnarray}
with
\begin{equation}
\label{eq: 4.3.5}
I(x,y)=\int_0^\infty {\rm d}t\ \frac{1}{t^2}\exp\left[-\left({\cal A}t+\frac{{\cal B}(x,y)}{t}\right)\right]
\end{equation}
where 
\begin{equation}
\label{eq: 4.3.5a}
{\cal A}=\frac{v_0^2}{4D_x}+\frac{v_y^2}{4D_y} 
,\ \ \ \ 
{\cal B}(x,y)=\frac{x^2}{4D_x} +\frac{y^2}{4D_y} \ .
\end{equation}
Using the Laplace method to approximate the integral $I(x,y)$, we find that the exponent in the integrand is minimised at time 
\begin{equation}
\label{eq: 4.3.6}
t^\ast=\sqrt{\frac{{\cal B}}{{\cal A}}}
\end{equation}
and the Laplace estimate for this integral is
\begin{equation}
\label{eq: 4.3.7}
I(x,y)=\sqrt{\frac{\pi}{2}}
\frac{{\cal A}^{1/4}}{{\cal B}^{3/4}}
\exp[-2\sqrt{\cal AB}] \ . 
\end{equation}
We are concerned with determining the leading order behaviour as $|X|\to \infty$. 
The Laplace approximation for $I(x,2X-\ln (2\cosh\,x)))$ is valid when ${\cal B}$ is sufficiently large, that is when $|X|$ is large. This is
\begin{eqnarray}
\label{eq: 4.3.7b}
\lefteqn {I(x,2X-\ln (2\cosh\, x))}  \nonumber \\
& & \!\!\!\!\!\!\!\!\!\!\!\!\!\!\!\!\!\!\!\!\!\!\!\!\!\!\!\!\!\!\!\!\!\!\!\!\!\
\sim \sqrt{\frac{\pi}{2}}{{\cal A}}^{1/4}\left(\frac{X^2}{D_y}\right)^{-3/4} 
\times \exp\left[-2\sqrt{{\cal A}}
\sqrt{\frac{X^2}{D_y}-\frac{X\ln(2\cosh\,x)}{D_y}
+\frac{[\ln(2\cosh\,x)]^2}{4D_y}+\frac{x^2}{4D_x}
}\right]
\nonumber \\
& & \!\!\!\!\!\!\!\!\!\!\!\!\!\!\!\!\!\!\!\!\!\!\!\!\!\!\!\!\!\!\!\!\!\!\!\!\!\!\!\
\sim \sqrt{\frac{\pi}{2}}{{\cal A}}^{1/4}D_y^{3/4}|X|^{-3/2}
\times\exp\left[-2\sqrt{{\cal A}}\frac{|X|}{\sqrt{D_y}}
\sqrt{1-\frac{\ln(2\cosh\,x)}{X}+O( X^{-2} ) }\right]
\nonumber \\
& & \!\!\!\!\!\!\!\!\!\
\sim {\cal C}|X|^{-3/2}
\times\exp\left[-2\sqrt{\frac{{\cal A}}{D_y}} 
\left(|X| + \frac{1}{2}\ln(2\cosh\,x)\right) \right]
\end{eqnarray}
where ${\cal C}$ is independent of $X$. 
From equations (\ref{eq: 4.3.4}) and (\ref{eq: 4.3.7}) we see that, to leading order 
as $X\to \infty$, the asymptotic behaviour of $P_X$ is 
\begin{equation}
\label{eq: 4.3.9}
P_X\sim |X|^{-3/2}\exp[-\Lambda |X|]
\ ,\ \ \ 
\Lambda \equiv 
\sqrt{\frac{v_0^2}{D_xD_y}+\frac{v_y^2}{D_y^2}} + \frac{v_y}{D_y} \ ,
\end{equation}
providing that the following integral is finite:
\begin{equation}
\label{eq: 4.3.10}
J=\int_0^\infty {\rm d}x\ x \exp\left[
\frac{v_0x}{2D_x}-\frac{v_y}{2D_y}\ln (2\cosh\,x)-\sqrt{\frac{{\cal A}}{D_y}\ln(2\cosh\,x)}
\right]
\ .
\end{equation}
This integral converges provided
\begin{equation}
\label{eq: 4.3.10a}
\sqrt{\frac{v_0^2}{D_xD_y}+\frac{v_y^2}{D_y^2}}-\frac{v_0}{D_y}+\frac{v_y}{D_y}>0
\ .
\end{equation}
This condition can be shown to be equivalent to the degeneracy condition, namely that at the stationary point $(Z_1^{\ast}, Z_2^{\ast})$ of $\Phi(Z_1,Z_2)$,  $Z_2^{\ast} > Z_1^{\ast} $. Because, in the $(x,y)$ coordinate system, ${\bf D}$ is diagonal and the function $\Phi(x,y)$ has the form
\begin{equation}
\label{eqn: 4.3.11}
\Phi(x,y)=\frac{1}{2} 
    \left[ \sqrt{ \frac{v_0^2}{D_x}+\frac{v_y^2}{D_y} }
       \sqrt{\frac{x^2}{D_x}+\frac{y^2}{D_y}}
-      \left( \frac{v_0}{D_x}+\frac{v_y}{D_y} \right)  
    \right] 
\end{equation}
Therefore, requiring that the minimum of $\Phi(x,y)$, on the line $y=2X-x$, is degenerate, and so lies at a negative value of $x$,  gives (\ref{eq: 4.3.10a}). Note that equation (\ref{eq: 4.3.9}) shows that in the degenerate case 
\begin{equation}
\label{eq: 4.3.12}
P_\epsilon \sim \left(\ln\,\frac{1}{\epsilon} \right)^\mu \epsilon^\gamma
\end{equation}
with $\mu=-3/2$, and $\gamma=\Lambda-1$.

\section{Advective flow model}
\label{sec: 5}

\subsection{Description of the model}
\label{sec: 5.1}

As a specific example, consider the matrix representing the differential of a random flow. Suppose that the velocity field of the flow, ${\bm u}\left({\bm x}(t),t\right)$, is defined by a random potential 
$\phi\left({\bm x}(t), t\right)$ and a random stream function $\mbox{\bm $\psi $}\left({\bm x}(t), t\right)$ 
such that
\begin{equation}
\label{eq: 5.1}
\mbox{\boldmath$u$}
=\mbox{\boldmath$\nabla$}\wedge \mbox{\boldmath$\psi$} +\beta \mbox{\boldmath$\nabla$}\phi \ ,
\end{equation}
where $\beta$  is a compressibility parameter. We assume that the fields 
$\mbox{\boldmath $\psi $}$ and $\phi$ are independent, have zero mean values and the same 
correlation function. We consider the case where the there are no temporal correlations, so that the  position of a particle advected by the fluid, 
${\bm x}(t)$, is a vector-valued random process satisfying
\begin{equation}
\label{eq: 5.2}
\mbox{\boldmath$x$}\left( t +\delta t \right) = 
        \mbox{\boldmath$x$} \left( t \right) +
        \mbox{\boldmath$u$} \left(\mbox{\boldmath$x$} \left( t \right), t\right)
				\sqrt{\delta t}
\end{equation}
and to leading order the separation $\delta \bm{r}=\bm{x}_1-\bm{x}_0$ of two nearby particles can be written as
\begin{equation}
\label{eq: 5.4}
    \delta \mbox{\boldmath$r$} \left( t +\delta t \right) \equiv
		\mbox{\boldmath$a$}(\mbox{\boldmath$x$}_0
		\left( t \right),t) \delta \mbox{\boldmath$r$} \left( t \right) =
    \left[  \mbox{\boldmath$I$} + \delta \mbox{\boldmath$a$} 
    \left(\mbox{\boldmath$x$}_0\left( t \right),t \right)  \right]
    \delta \mbox{\boldmath$r$} \left( t \right)
\end{equation}
where
\begin{equation}
\label{eq: 5.5}
  \left[ \delta \mbox{\boldmath$a$} \left(\mbox{\boldmath$x$}_0\left( t \right),t \right) \right]_{ij} =
   \left(\frac{\partial u_i \left(\mbox{\boldmath$x$}_{0}(t), t\right)}{\partial x_j} \right) \sqrt{\delta t}.
\end{equation} 
The elements of the matrix $ \delta {\bm a}$  are random variables constructed from the second derivatives of the velocity field potentials evaluated at time $t$ and position ${\bm x}_0(t)$ :
\begin{equation}
\label{eq: 5.6}
\delta {\bf a}=\left(\begin{array}{cc}
\psi_{xy}+\beta \phi_{xx} & -\psi_{yy}+\beta \phi_{xy}\cr 
\psi_{xx}+\beta \phi_{xy} & -\psi_{xy}+\beta \phi_{yy}
\end{array}\right)\sqrt{\delta t}.
\end{equation}
All the derivatives have mean value zero, and are normalized so that their non-zero covariances are:
\begin{equation}
\label{eq: 5.7a}
\langle \psi_{xx}^2\rangle = \langle\psi_{yy}^2\rangle =3 
\ \ , \ \ \ \ 
\langle \psi_{xx}\psi_{yy}\rangle = \langle \psi_{xy}^2\rangle =1\ ,  
\end{equation}
and similarly for derivatives of $\phi$. Also, since the model is rotationally invariant, and since $\delta \tilde{{\bf a}}_{n}$ , defined by (\ref{eq: 2.1x}) is a rotational transformation of $\delta{\bf a}_{n}$ the elements of $\delta \tilde{{\bf a}}_{n}$ have the same statistics as those of $\delta{\bf a}_{n}$.

From equations (\ref{eq: 5.6}) and (\ref{eq: 5.7a}), using equations (\ref{eq: 2.7}), (\ref{eq: 2.8}) and (\ref{eq: 2.11}), it follows that the drift velocity and diffusion tensor for this model are, respectively:
\begin{equation}
\label{eq: 5.7b}
{\bm v}=\left(\begin{array}{cc}
1-\beta^2 \cr 
-(1+3\beta^2)
\end{array}\right)
\end{equation}
\begin{equation}
\label{eq: 5.7c}
{\bf D}=\frac{1}{2}\left(\begin{array}{cc}
1+3\beta^2,    &    \beta^2 - 1 \cr 
\beta^2 - 1,   &     1+3\beta^2
\end{array}\right).
\end{equation}

\subsection{Theoretical predictions for $\gamma$} 
\label{sec: 5.2}

Using the above expressions for ${\bm v}$ and ${\bf D}$ to compute the 
function $\Phi(Z_1,Z_2)$ using equation (\ref{eq: 3.4}) gives
\begin{equation}
\label{eq: 5.10}
\!\!\!\!\!\!\!\!\!\!\!\!\!\!\!\!\!\!\!\!\
\Phi(Z_1,Z_2)=\frac{1}{2\beta}\sqrt{\frac{1+3\beta^2}{2(1+\beta^2)}}
\sqrt{(1+3\beta^2)(Z_1^2+Z_2^2)+2(1-\beta^2)Z_1Z_2}+Z_2
\ .
\end{equation}
Considering the discussion immediately preceding equation (\ref{eq: 3.6}), the critical point is 
\begin{equation}
\label{eq: 5.10a}
{\bm Z}^\ast=|X|{\bm u}^\ast=|X|\left(\begin{array}{cc} -1 \cr -\eta^\ast \end{array}\right)
\end{equation}
where $f^{\prime}(\eta^\ast)=0$ with $f(\eta) \equiv \Phi(-1,-\eta)$. Therefore, in terms of the compressibility parameter, $\beta$, we have 
\begin{equation}
\label{eq: 5.10b}
\eta^\ast=\frac{7 \beta^4+10\beta^2-1}{(1+3\beta^2)(1-\beta^2)} \ .
\end{equation}
Noting that 
\begin{equation}
\label{eq: 5.10c}
(\eta^\ast-1)=\frac{2(1+\beta^2)(5 \beta^2-1)}{(1+3\beta^2)(1-\beta^2)} 
\end{equation}
if $1 > \beta > 1/\sqrt{5}$ then we must have $\eta^\ast > 1$ so that $Z_2^\ast < X=Z_1^\ast$ and the critical point is non-degenerate. However, if $0 \le \beta < 1/\sqrt{5}$ then $\eta^\ast < 1$ so that 
$Z_2^\ast > X=Z_1^\ast$, in which case the critical point is degenerate.

In the non-degenerate case, we find that $\mbox{$\Phi({\bm u}^\ast)=f(\eta^\ast)=
-2(\beta^2-1)/(1+3\beta^2)$}$ and, from equation (\ref{eq: 3.7}),
\begin{equation}
\label{eq: 5.10d}
\gamma= -\frac{2(1+\beta^2)}{(1+3\beta^2)} -1\ .
\end{equation}
In the degenerate case, setting $\eta^\ast=1$ so that $Z_1^\ast=Z_2^\ast=X$, and noting that, since $X<0$,  $\sqrt{Z_1^\ast Z_2^\ast}=|X|=-X$, gives 
$\Phi({\bm u}^\ast)=\sqrt{(1+3\beta^2)/(2\beta^2)}-1$ so that
\begin{equation}
\label{eq: 5.10e}
\gamma= \frac{1}{\beta} \sqrt{\frac{1+3\beta^2}{2}}-2
\ .
\end{equation}
This expression is suspect, however, because the derivation of (\ref{eq: 3.7}) 
depends on the assumption that $Z_1>Z_2$, which is violated by the critical point 
condition. 

\subsection{Exact equations for evolution of singular values}
\label{sec: 5.3}

In order to understand the degenerate case in more detail, we need to consider a more refined treatment that does not assume $Z_1\gg Z_2$.
Using the statistics for the increments ${\rm d}\tilde{a}_{ij}$, obtained from (\ref{eq: 5.6}) and (\ref{eq: 5.7a}), the exact equations of motion in the the $(x,y)$ coordinate system, equations (\ref{eq: 4.1.1}), become:
\begin{eqnarray}
\label{eq: 5.12}
{\rm d}x&=&\left({\rm d}\tilde{a}_{11}-{\rm d}\tilde{a}_{22}\right)
+2(1+\beta^2)\frac{1+\nu^2}{1-\nu^2}{\rm d}t
\nonumber \\
{\rm d}y&=&\left({\rm d}\tilde{a}_{11}+{\rm d}\tilde{a}_{22}\right)
-4\beta^2{\rm d}t
\end{eqnarray}
and the second moments of the increments are therefore
\begin{equation}
\label{eq: 5.13}
\langle {\rm d}x^2\rangle = 4(1+\beta^2){\rm d}t, \ \ \
\langle {\rm d}y^2\rangle = 8\beta^2{\rm d}t, \ \ \
\langle {\rm d}x{\rm d}y\rangle = 0 \ .
\end{equation}
Noting that $\nu=\lambda_2/\lambda_1=\exp(-x)$ we see that $x$ and $y$ make independent diffusive motions, with the following drift velocity and diffusion tensor:
\begin{equation}
\label{eq: 5.18}
{\bm v}=\left(\begin{array}{cc}
2(1+\beta^2){\rm coth}(x)\cr 
-4\beta^2
\end{array}\right)
\ , \ \ \ 
{\bf D}=\left(\begin{array}{cc}
2(1+\beta^2)    &    0 \cr 
0   &     4\beta^2
\end{array}\right).
\end{equation}
These results are consistent with the remarks regarding $v_y$ and $v_x$ in sections (\ref{sec: 4.1}) and (\ref{sec: 4.2}), respectively: $v_y$ is constant and, since $v_x(x)=2(1+\beta^2) {\rm coth}(x)$, we have 
$v_0=2(1+\beta^2)$ and $v_x \to \infty$ as $x \to 0$.

\section{Ratio of singular values}
\label{sec: 6}

Thus far we have considered the PDF of the norm, $\epsilon$, of the matrix product in the limit as $\epsilon \to 0$, showing that it is a power-law, with an additional logarithmic correction in the degenerate case. We can also consider conditional probabilities. One interesting example is the distribution of the ratio of singular values $\lambda_2/\lambda_1$, for a given value of $\epsilon$. Equivalently, we can consider the PDF of $x =\Delta Z = Z_1-Z_2 = \ln (\lambda_1/\lambda_2)$, for a given value of $X =\ln\,\epsilon$, this will be denoted $P_{\Delta Z\vert X}$.

In the non-degenerate case it is clear from the discussion in section \ref{sec: 3} that the ratio $\lambda_2/\lambda_1$ approaches zero as $\epsilon\to 0$. In the degenerate case, however, the values of $Z_1$ and $Z_2$ are comparable and the distribution of $\Delta Z$ is non-trivial. Accordingly, we concentrate on $P_{\Delta Z\vert X}$ for the degenerate case. To simplify the discussion we only give explicit formulae for the case where $\vert \Delta Z/X \vert \ll 1$.

The joint distribution of $\mbox{\boldmath$Z$}=(Z_1,Z_2)$ takes the form given by equation (\ref{eq: 3.11}), namely 
$P_{\bm Z}(\mbox{\boldmath$Z$})
\sim (\mbox{\boldmath$Z$}\cdot{\bf D}^{-1}\mbox{\boldmath$Z$})^{-1/4}
\exp[-\Phi(\mbox{\boldmath$Z$})]$. 
We are interested in the distribution 
\begin{equation}
\label{eq: 6.2}
P_{\Delta Z\vert X}=\frac{P_{\bm Z}(X,X-\Delta Z)}{\int_0^\infty {\rm d}\Delta Z\ P_{\bm Z}(X,X-\Delta Z)} \ .
\end{equation}
Provided $\vert \Delta Z/X \vert \ll 1$ the dependence of the pre-exponential factor in (\ref{eq: 3.11}) on $\Delta Z$ can be neglected and we have 
\begin{equation}
\label{eq: 6.3}
P_{\Delta Z\vert X}={\cal K}\exp[\Phi(X,X)-\Phi(X,X-\Delta Z)]
\end{equation}
where ${\cal K}$ is a normalisation factor.
Equation (\ref{eq: 6.3}) is based upon the assumption that the $Z_i$ undergo diffusion with a constant drift velocity. This assumption ceases to be valid close to the reflecting boundary, $x=0$ where $\Delta Z$ is very small and the drift velocity is given by (\ref{eq: 5.18}). 

We are interested in finding a solution which matches (\ref{eq: 4.2.15}) when $x \gg 1$ but which obeys the correct Fokker-Planck equation, namely
\begin{equation}
\label{eq: 6.4}
\frac{1}{2(1+\beta^2)}\partial_t P=\partial_x^2 P -\partial_x\left[\frac{1}{{\rm tanh}(x)}P\right]
\end{equation}
Making the same transformation to Hermitean form as equations (\ref{eq: 4.2.2})-(\ref{eq: 4.2.4}), and setting $\psi(x,t)=\phi(x)\exp[-v_0^2t/4D_x]$ in order to find a solution which matches (\ref{eq: 4.2.15}), we find that $\phi(x)$ satisfies:
\begin{equation}
\label{eq: 6.8}
\phi ^{\prime \prime} =-\frac{1}{4\,{\rm sinh}^2(x)}\phi
\end{equation}
We require a solution $\phi(x)$ which approaches a constant as $x\to \infty$ and which approaches zero as $x\to 0$. Close to $x=0$ the differential equation is approximated by  $\phi ^{\prime \prime} =  -\phi/4x^2 $. This has general solution $\phi(x)=\sqrt{x} (a+b \ln(x) )$, where $a$ and $b$ are arbitrary constants. Since $\phi(x) \to 0$ as $x \to 0$  we conclude that $ \phi(x) \sim \sqrt{x}$ as $x \to 0$.

From equation (\ref{eq: 4.2.2}) with $v_x=D_x \coth(x)$ we have $\chi(x) = \ln(\sqrt{\sinh(x)})$ and, therefore, $\exp(\chi(x) ) \to \sqrt{x}$ as $x \to 0$. 
Hence, from equations (\ref{eq: 4.2.2}) and (\ref{eq: 6.2}) we conclude that $P_{\Delta Z\vert X}$ is of the form 
\begin{equation}
\label{eq: 6.9}
P_{\Delta Z\vert X}=F(\Delta Z)\exp[\Phi(X,X)-\Phi(X,X-\Delta Z)]
\end{equation}
where $F(\Delta Z)\sim \Delta Z$ for $\Delta Z\ll 1$, but where $F(\Delta Z)$ approaches a constant as $\Delta Z\to \infty$.

\section{Numerical investigations  of the advective flow model}
\label{sec: 7}

Our analysis of the matrix contraction process has led us to consider a diffusive model for the evolution of the singular values. The predictions of this model
are a consequence of the fact that the diffusion process has an unusual combination of reflecting and absorbing boundary conditions. Because the problem is too complex for a rigorous analysis to be practicable, we have tested the predictions by means of numerical simulations. 

\begin{figure*}[!htbp]
\label{fig: 2}
\begin{center}
\subfigure[]{ \includegraphics[width=0.475\textwidth]{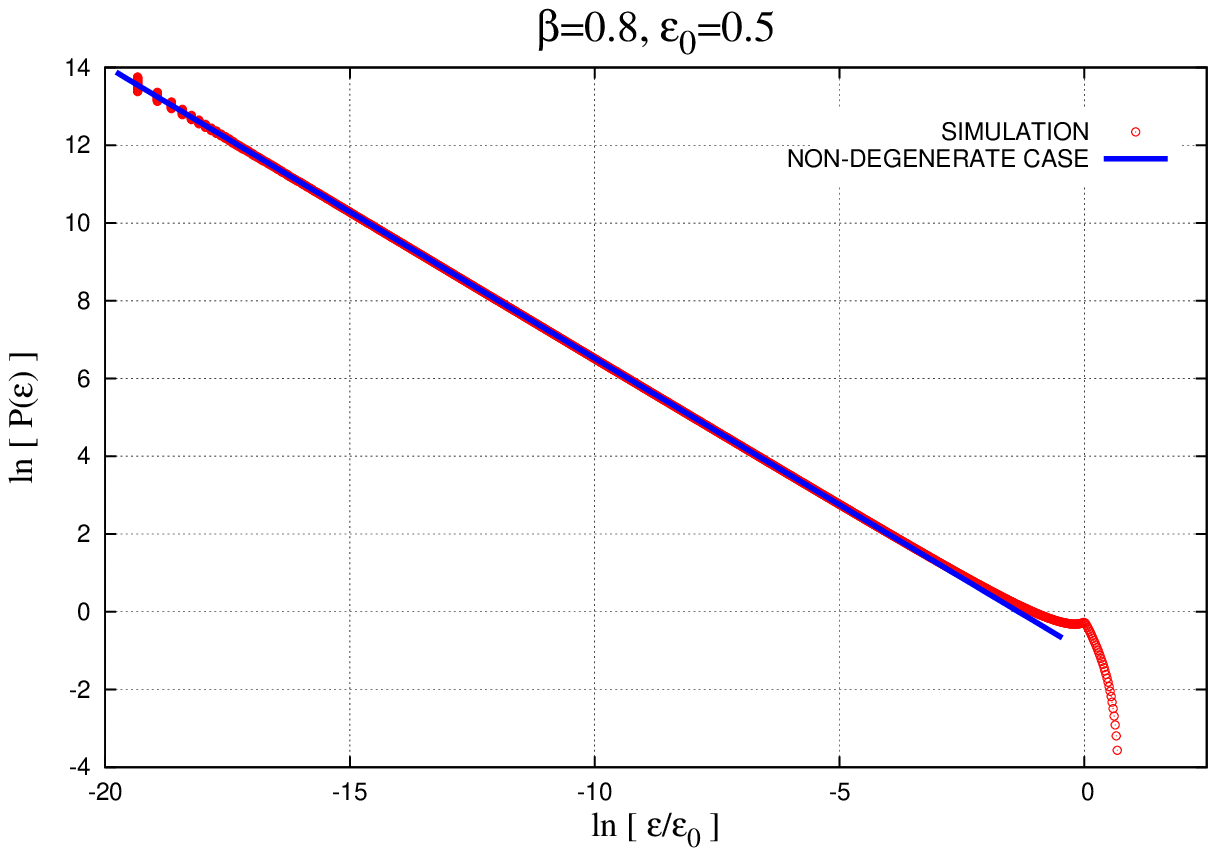} }
\subfigure[]{ \includegraphics[width=0.475\textwidth]{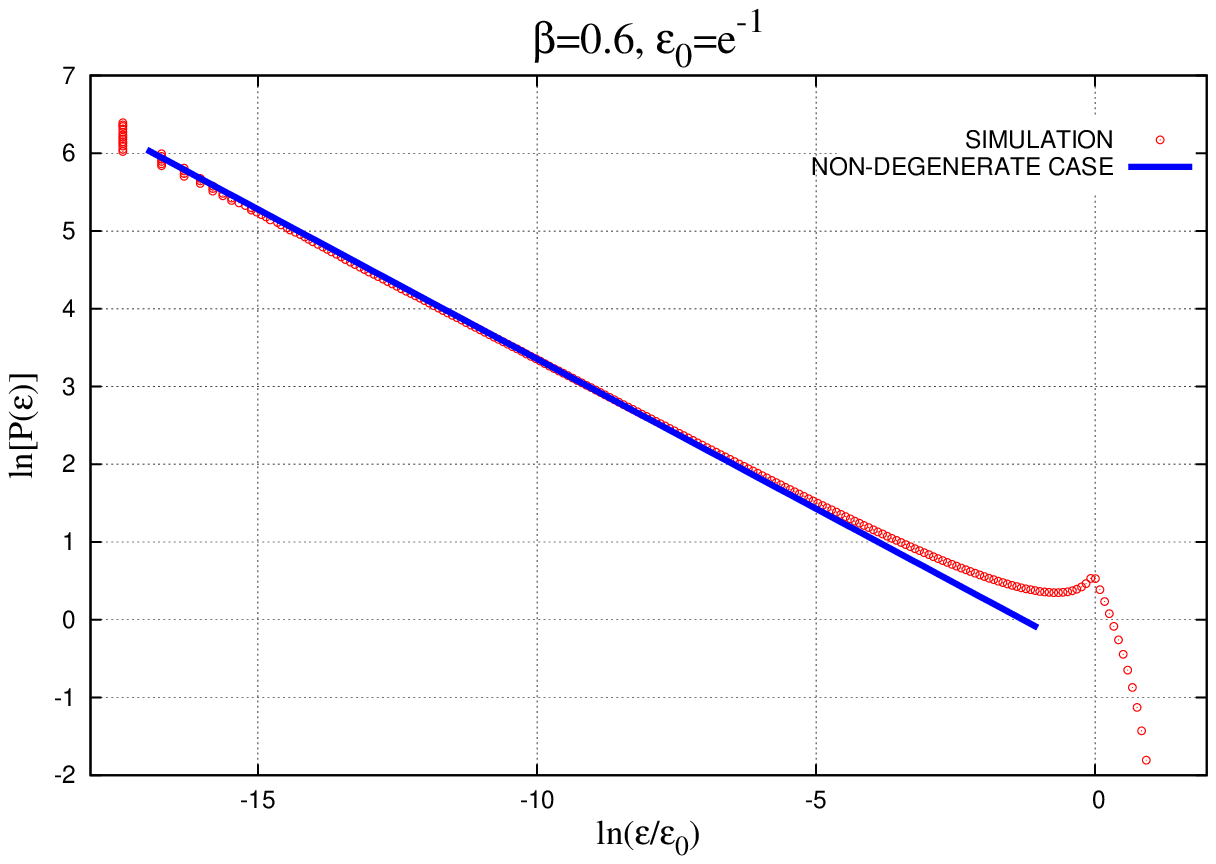} }
\subfigure[]{ \includegraphics[width=0.475\textwidth]{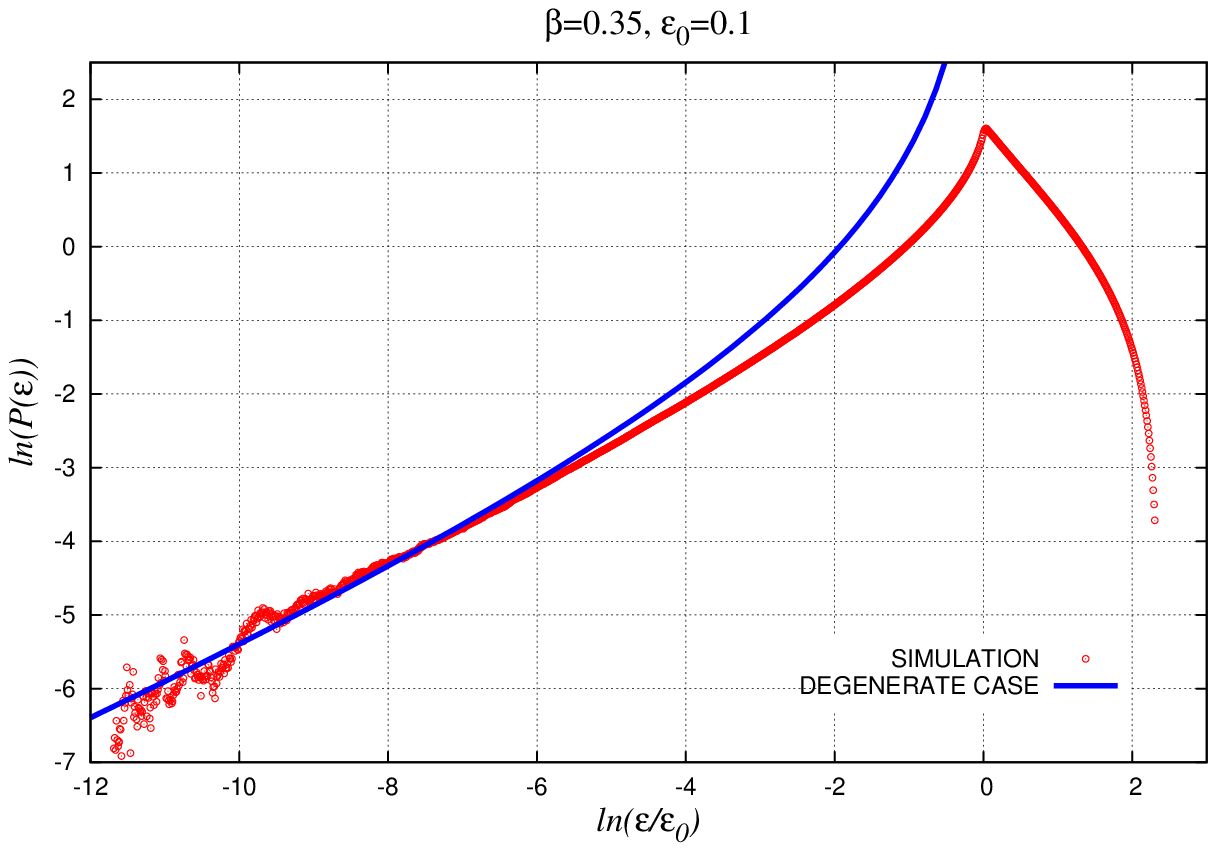} }
\subfigure[]{ \includegraphics[width=0.475\textwidth]{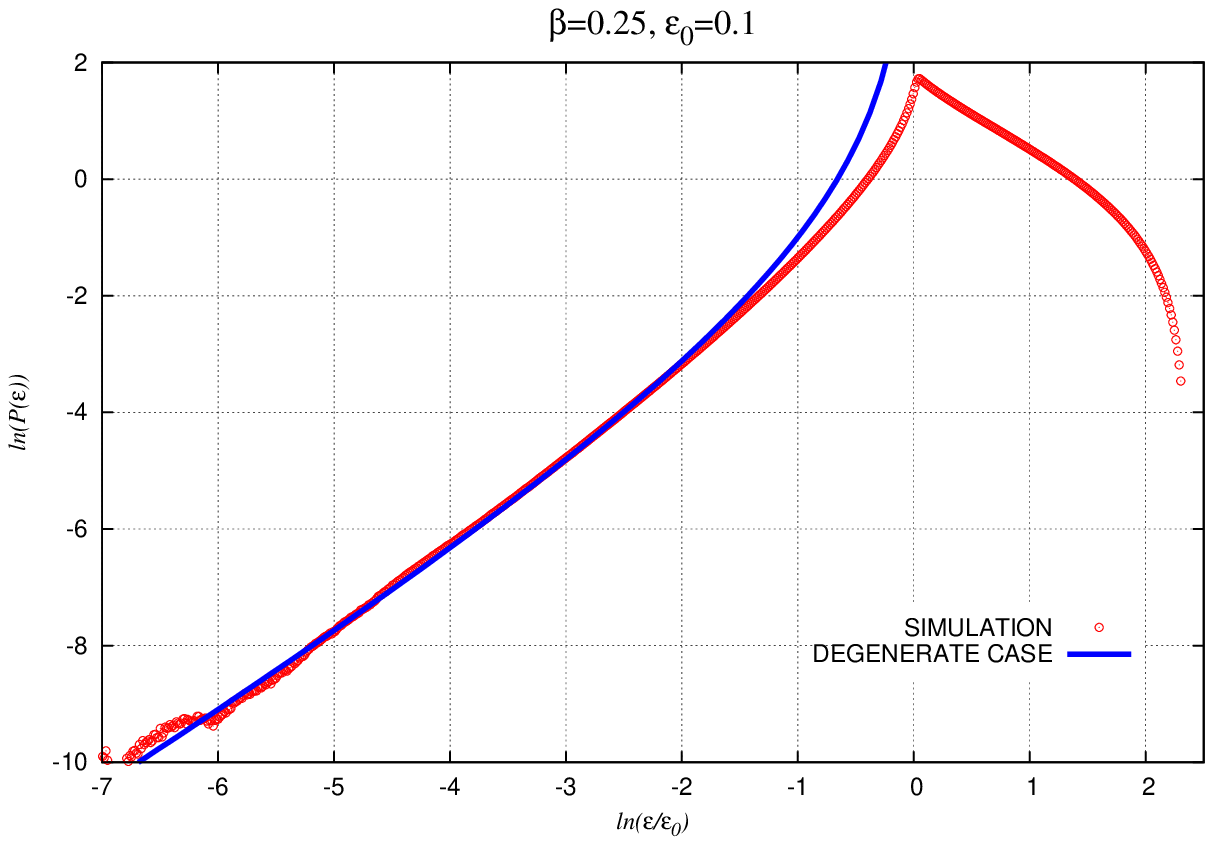} }
\end{center}
\caption{(Colour online) Simulation of the PDF of the matrix norm for the matrix contraction process, where the matrices are stability matrices for the random advection model discussed in section \ref{sec: 5}. The simulated distributions are plotted in red, the theoretical distributions are in blue.
For ({\bf a}), $\beta=0.8$, and ({\bf b}), $\beta=0.6$, the stationary point is non-degenerate, and the distribution is a simple power-law, with equation (\ref{eq: 5.10d}) predicting $\gamma=-0.753$ and $\gamma=-0.385$ respectively. For ({\bf c}), $\beta=0.35$, and ({\bf d}), $\beta=0.25$, the 
stationary point is degenerate and the PDF of $\epsilon$ has a factor $\left(\ln\,\frac{1}{\epsilon}\right)^{-3/2}$. Equation (\ref{eq: 5.10e}) predicts that $\gamma=0.363$ and $\gamma=1.082$
respectively.} 
\end{figure*}  

The results of direct simulations of the matrix contraction process, using the random advection model discussed in section \ref{sec: 5} to generate the ensemble of random matrices, are given in plots (a) and (b) of figure 2. In plots (c) and (d), which deal with the degenerate case,  we show the results obtained using the exact equations of motion for the singular values, as given in section (\ref{sec: 5.3})  (having first verified that these equations produce identical results to the direct simulation approach).
These results show that the predictions for $\gamma$ are correct for both the non-degenerate 
case ($1 > \beta > 1/\sqrt{5}$) and the degenerate case ($0 <  \beta < 1/\sqrt{5}$).

\begin{figure*}[!htbp]
\begin{center}
\includegraphics[angle=0, width=0.65\textwidth]{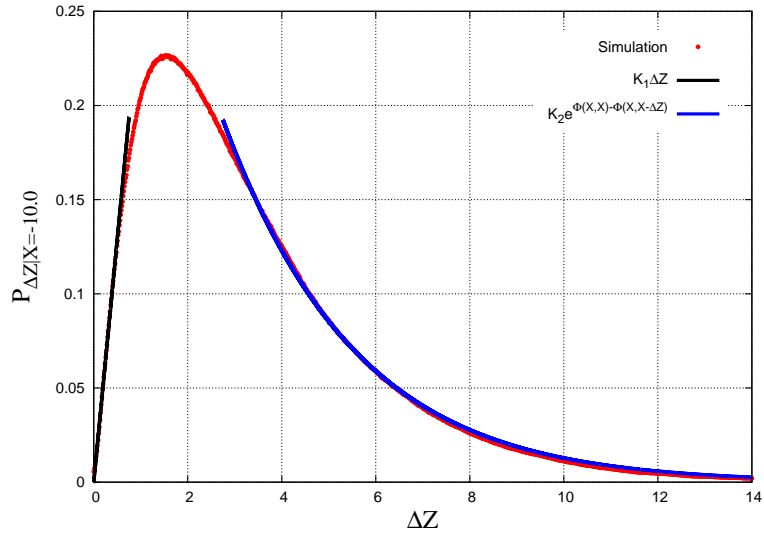} 
\end{center}
\caption{ (Colour online). The result of a simulation of the conditional PDF $P_{{\Delta Z} | {\rm X}} $, for the degenerate case where $\beta =0.3$, with $\epsilon_0=0.01$ and ${\rm X}=-10$. The result is consistent with equation (\ref{eq: 6.9}): for $\vert \Delta Z/X \vert \ll 1$  $P_{{\Delta Z} \vert {\rm X}} \sim {\Delta Z}$ and for $\vert \Delta Z/X \vert \gg 1$ the distribution has an exponential tail.}
\label{fig: 7.3}
\end{figure*}

Figure \ref{fig: 7.3} shows the result of a simulation of the conditional PDF $P_{{\Delta Z} \vert {\rm X}}$, where $\Delta Z=Z_1-Z_2 $ and $ {\rm X}= {\rm ln} (\epsilon)$, for the degenerate case 
where $\beta=0.3$, with $\epsilon_0=0.01$ and ${\rm X} = -10$. The plot also shows segments of two fitted curves: the straight line $ P_{{\Delta Z} \vert {\rm X}}=K_1 {\Delta Z}$ where $K_1$ is a constant (shown in black) and the exponential tail  
$P_{{\Delta Z} \vert {\rm X}}=K_2 \exp{[\Phi(X,X)-\Phi(X,X-{\Delta Z})]}$ where $K_2$ is a constant (shown in blue). These demonstrate that the form of $P_{{\Delta Z}|{\rm X}}$ is as given in equation (\ref{eq: 6.9}).
We remark that it is not actually necessary to compute the velocity field to simulate the 
matrices $\delta {\bf a}$.  Defining the vectors 
${\bm C}  = (C_1,C_2,C_3)=(\psi_{xx},\psi_{yy},\psi_{xy})$ and 
${\bm D}=(D_1,D_2,D_3)=(\phi_{xx},\phi_{yy},\phi_{xy})$ 
then the covariance matrix of ${\bm C}$ and of ${\bm D}$ is
\begin{equation}
\label{eq: 7.2.1}
{\bm K}=\langle C_iC_j\rangle=
\left(\begin{array}{ccc}
3 & 1 & 0 \cr
1 & 3 & 0 \cr
0 & 0 & 1 \cr
\end{array}\right) .
\end{equation}
Therefore if ${\bm \xi}= (\xi_1,\xi_2,\xi_3)$ and ${\bm \eta}=(\eta_1,\eta_2,\eta_3)$  are each vectors whose elements are uncorrelated Gaussian random variables with zero mean and unit variance, we may write ${\bm C} ={\bm K}^{1/2} {\bm \xi}$, and ${\bm D} ={\bm K}^{1/2} {\bm \eta} $, where
\begin{equation}
\label{eq: 7.2.2}
{\bf K}^{1/2}=
\frac{1}{\sqrt{2}}
\left(\begin{array}{ccc}
\sqrt{2}+1 & \sqrt{2}-1 & 0 \cr
\sqrt{2}-1 & \sqrt{2}+1 & 0 \cr
0 & 0 & \sqrt{2}
\end{array}\right)
\end{equation}
We used this approach to simulate the stability matrices of the random flow model.

\section{Conclusions}
\label{sec: 8}

We have investigated a process which occurs naturally in models for chaotic 
dynamical systems, considering the distribution of small values of the norm of the 
product of random matrices, and resetting the process 
to the identity matrix whenever the norm ceases to be small. We considered a model 
involving a product of matrices which are close to the identity, and which have diffusive fluctuations. For the scalar version of this problem, the distribution of $\epsilon$ is always a power-law. For the matrix contraction process, we find that the distribution is of the form (\ref{eq: 1.5}) when the matrix has diffusive fluctuations, with two possible values of $\mu$ ($0$ or $-3/2$). 
It would be of interest to know about the distribution of $\epsilon$ for more general 
classes of matrix. 

{\sl Acknowledgements}.  We are grateful for the hospitality of the Department 
of Physics at the University of Auckland, where much of this paper was written.


\section*{References}
\begin{thebibliography}{10}

\bibitem{Gui+16}
R. Guichardaz, A. Pumir and M. Wilkinson,
{\it Europhys. Lett.}, {\bf 115}, 10009, (2016).
%
\bibitem{Ott02}
E. Ott,
{\sl Chaos in Dynamical Systems}, 2nd edition, Cambridge: University Press, (2002).
%
\bibitem{Wil+17}
M. Wilkinson and J. Grant,
{\sl Statistics of Contracted Constellations of a Dynamical System}, 
in preparation, for submission to {\it J. Phys. A.}
%
\bibitem{Horn13}
R. A. Horn and C. R. Johnson,
{\em Matrix Analysis }, 2nd. edn., Cambridge University Press, New York, (2013)..
%
\bibitem{Gra+15}
J. Grant and M. Wilkinson,
{\it J. Stat. Phys.}, {\bf 160}, 622-35, (2015).
%
\bibitem{Wil+15}
M. Wilkinson, R. Guichardaz, M. Pradas and A. Pumir, 
{\it Europhys. Lett.}, {\bf 111}, 50005, (2015).
%
\bibitem{Wil+03}
M. Wilkinson and B. Mehlig,
{\it Phys. Rev. E}, {\bf 68}, 040101, (2003).
%
\bibitem{Wil+12}
M. Wilkinson, B. Mehlig, K. Gustavsson and E. Werner,
{\it Eur. Phys. J. B}, {\bf 85}, 18, (2012).
%
\bibitem{Som+93}
J. Sommerer and E. Ott,
{\it Science}, {\bf 359}, 334, (1993).
%
\bibitem{Lar+09}
J. Larkin, M. M. Bandi, A. Pumir and W. I. Goldburg,
{\it Phys. Rev. E}, {\bf 80}, 066301,( 2009).

\bibitem{vKa81}
 N. G. van Kampen,
{\sl Stochastic processes in Physics and Chemistry}, 2nd ed.,
North-Holland, Amsterdam, (1981).
%
\bibitem{Abr+72}
M. Abramowitz and I. A. Stegun (eds.),
{\sl Handbook of Mathematical Functions},
New York: Dover, (1972).
%
\bibitem{Fal+01}
G. Falkovich, K. Gawedzki and M. Vergassola,
{\it Rev. Mod. Physics }, {\bf 73}, 913-975, (2000).

\bibitem{Bec+04}
J. Bec, K. Gawedzki and P. Horvai,
{\it Phys. Rev. Lett.}, {\bf 92}, 224501, (2004).

\bibitem{Mah09}
 R. Mahnke, J. Kaupu$\check{\rm{z}}s$ and I. Lubashevsky,
{\sl Physics of Stochastic Processes}, 2nd ed.,
Wiley-VCH, Weinheim, (2009).
%

\end{thebibliography}
\end{document}